# Chiral Structured Illumination Microscopy

Shiang-Yu Huang,[1] Jiwei Zhang,[1] Christian Karras,[1] Ronny Förster,[1]
Rainer Heintzmann[1,2,3,*] and Jer-Shing Huang[1,2,4,5,*]

[1]*Leibniz Institute of Photonic Technology, Albert-Einstein Straße 9, 07745 Jena, Germany*
[2]*Abbe Center of Photonics, Friedrich-Schiller University Jena, Jena, Germany*
[3]*Institute of Physical Chemistry, Friedrich-Schiller University Jena, Jena, Germany*
[4]*Research Center for Applied Sciences, Academia Sinica, 128 Sec. 2, Academia Road, Nankang District, 11529 Taipei, Taiwan*
[5]*Department of Electrophysics, National Chiao Tung University, 1001 University Road, 30010 Hsinchu, Taiwan*

*e-mail address: heintzmann@gmail.com; jer-shing.huang@leibniz-ipht.de*

We propose a chiral imaging modality based on optical chirality engineering, fluorescence-detected circular dichroism and structured illumination microscopy. In this method, the optical chirality of the illumination is structured and the circular dichroism dependent fluorescence is detected. With image reconstruction, the spatial distribution of chiral domains can be obtained at sub-diffraction limited resolution. We theoretically demonstrate this method and discuss the feasibility using an optical chirality engineering approach based on far-field optics.

Chiral objects with opposite handedness interact differently with circularly polarized light (CPL), leading to circular birefringence and circular dichroism (CD). These chiroptical responses provide rich structural information of the chiral objects [1] and have found important applications in biomedical, pharmaceutical and material sciences [2]. CD spectroscopy measures the wavelength-dependent difference in absorption of chiral light by chiral objects and has been applied to obtain structural information of important biological targets, like proteins [3-6]. However, CD spectroscopy does not provide spatial information of the molecular chiral domains, which is essential in two-dimensional dynamic analysis or *in vivo* imaging. To obtain the spatial distribution of chiral domains, there have been attempts to perform direct chiral imaging. Currently existing chiral imaging methods include direct wide-field CD imaging [7-9], confocal CD mapping [10,11], second harmonic generation CD (SHG-CD) mapping [12-18], two-photon luminescence (TPL) chirality mapping [19], CD thermal lensing (CD-TL) microscopy [20,21] and chiral near-field scanning optical microscopy (NSOM) [22-24]. Main limitations of these methods are low throughput, weak contrast and diffraction-limited resolution. For example, the image contrast in direct wide-field CD imaging and confocal CD mapping is inherently low because the CD signal is weak [25]. SHG-CD mapping offers good spatial resolution and selectivity but can only be applied to samples containing ordered mesoscopic structures. CD-TL microscopy has unsatisfactory signal-to-noise ratio and requires long acquisition time with a lock-in amplifier. Confocal CD mapping, SHG-CD mapping and TPL chirality mapping have limited throughput due to their scanning nature. Chiral NSOM offers sub-wavelength resolution. However, its image acquisition rate is relatively slow and limited to the sample surface. Thus, a high-throughput wide-field imaging modality for chiral domains with sub-wavelength spatial resolution is highly desired.

In this letter, we propose a chiral imaging method which combines optical chirality (OC) engineering, fluorescence-detected circular dichroism (FDCD) [26-28] and structured illumination microscopy (SIM) [29-32]. This new imaging method (called chiral SIM) is based on wide-field illumination with fluorescence detection and provides sub-diffraction limited resolution. In typical SIM, structured intensity patterns are used for illumination (Fig. 1(a)). This leads to the moiré effect, which down-modulates the high spatial frequency information of the fluorescent sample into the detectable range of a microscope and thereby improves the spatial resolution after image reconstruction. In chiral SIM, the OC of the illumination is spatially structured (Fig. 1(b)). This generates a moiré pattern on fluorescent chiral domains and enables sub-diffraction limited chiral imaging.

The absorption rate of a chiral molecule illuminated by an external time-harmonic electromagnetic field can be expressed as [25]

$$A = \frac{\omega}{2}\alpha''|\mathbf{E}|^2 + \omega G''\text{Im}(\mathbf{E}^* \cdot \mathbf{B}), \qquad (1)$$

where $\omega$ is the angular frequency, $\alpha''$ is the imaginary electric dipole polarizability and $G''$ is the imaginary electric-magnetic mixed polarizability. $\mathbf{E}$ and $\mathbf{B}$ are the time-harmonic complex local electric and magnetic field, respectively. The contributions involving the magnetic susceptibility and higher-order moments are neglected as they are insignificant for typical molecules. Taking the optical chirality $C = -\dfrac{\varepsilon_0 \omega}{2}\operatorname{Im}(\mathbf{E}^* \cdot \mathbf{B})$ and electric energy density $U_e = \dfrac{\varepsilon_0}{4}|\mathbf{E}|^2$ where $\varepsilon_0$ is the permittivity of free space, Eq. (1) can be rewritten as [25]

$$A = \frac{2}{\varepsilon_0}\left(\omega U_e \alpha'' - C G''\right). \tag{2}$$

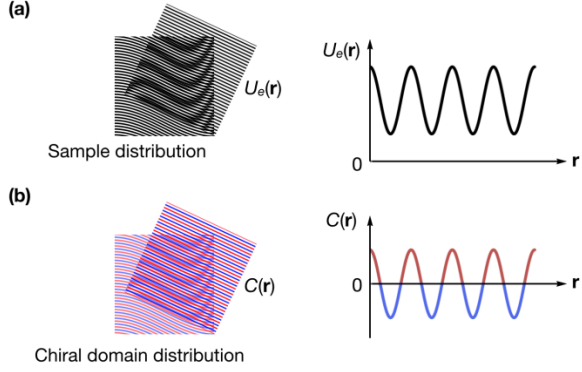

FIG. 1. (a) Left: moiré pattern generated by the superposition of sample distribution $\alpha''(\mathbf{r})$ and structured electric energy density $U_e(\mathbf{r})$ in typical SIM. Right: spatial profile of $U_e(\mathbf{r})$. (b) Left: moiré pattern generated by the superposition of chiral domain distribution $G''(\mathbf{r})$ and structured OC pattern $C(\mathbf{r})$ in chiral SIM. Right: spatial profile of $C(\mathbf{r})$. Red and blue colors denote the left and right handedness.

The first term on the right-hand side of Eq. (2) is the dominant electric dipole absorption and the second term is the chirality-dependent absorption, i.e., the origin of CD. The contribution of the second term requires the molecule to be chiral ($G'' \neq 0$) and the illumination to possess non-zero OC ($C \neq 0$). Depending on the OC of the illumination, the enantiomers of a chiral molecule show slightly different absorption, resulting in the chiral contrast. Since $\alpha''$ and $G''$ are determined by the molecular conformation, the way to externally modulate the absorption is to control the $U_e$ or $C$ of the illumination [25,33-37]. This possibility to modulate the absorption rate by engineering the OC of the illumination lays the foundation of the proposed chiral SIM method.

In conventional CD imaging [7,8], chiral samples are successively illuminated by left- ($+$) and right-handed ($-$) CPL with uniform intensity and spatially invariant OC ($C_\pm(\mathbf{r}) = \pm\dfrac{\varepsilon_0 \omega}{2c}|\mathbf{E}|^2$). The spatial distribution of absorption difference is therefore expressed as $\Delta A(\mathbf{r}) = A_+(\mathbf{r}) - A_-(\mathbf{r}) = -\dfrac{2\omega}{c}|\mathbf{E}|^2 G''(\mathbf{r})$. However, the spatial resolution of the CD distribution is fundamentally diffraction-limited. To improve the resolution, chiral SIM modulates the OC of the illumination and detects the CD-dependent incoherent fluorescence, which is essential to obtain sub-diffraction limited resolution [38]. Figure 1(b) illustrates the moiré pattern formed by the superposition of the structured OC pattern and the chiral domain distribution. To image chiral domains via fluorescence, the fluorophores have to comply with the criteria for FDCD and show CD-dependent fluorescence intensity [26]. This means the fluorophores should either be chiral or attached to the chiral domains, and the rotatory Brownian motion should randomize the orientation during the excited state lifetime of the fluorophores [27,28]. With these conditions satisfied, the fluorescence dependent on CD is

$$F(\mathbf{r}) = \beta A(\mathbf{r}), \tag{3}$$

where $\beta$ is a proportional constant that accounts for the quantum yield of the fluorophore and the detection efficiency of the optical system. Considering the point spread function $h(\mathbf{r})$ of the optical system, the acquired CD-dependent fluorescent image is

$$\begin{aligned}M(\mathbf{r}) &= F(\mathbf{r}) \otimes h(\mathbf{r}) \\ &= \frac{2\beta}{\varepsilon_0}\left[\omega U_e(\mathbf{r})\alpha''(\mathbf{r}) - C(\mathbf{r})G''(\mathbf{r})\right] \otimes h(\mathbf{r}),\end{aligned} \tag{4}$$

where $\otimes$ denotes the convolution operation. In chiral SIM, the OC is structured in a cosinusoidal form of $C(\mathbf{r}) = C_0 \cos(\mathbf{k}_C \cdot \mathbf{r} + \theta)$, where $C_0$ is the amplitude, $\mathbf{k}_C$ the $k$-vector and $\theta$ the phase of the structured OC. The Fourier transform of the acquired CD-dependent fluorescent image is therefore given by

$$\begin{aligned}\tilde{M}(\mathbf{k}) &= \frac{2\beta}{\varepsilon_0}\left[\omega \tilde{U}_e(\mathbf{k})\tilde{\alpha}''(\mathbf{k}) - \tilde{C}(\mathbf{k}) \otimes \tilde{G}''(\mathbf{k})\right]\tilde{h}(\mathbf{k}) \\ &= \frac{2\beta}{\varepsilon_0}\Big[\omega \tilde{U}_e(\mathbf{k})\tilde{\alpha}''(\mathbf{k}) - C_0 e^{-i\theta}\tilde{G}''(\mathbf{k}+\mathbf{k}_C) \\ &\quad - C_0 e^{+i\theta}\tilde{G}''(\mathbf{k}-\mathbf{k}_C)\Big]\tilde{h}(\mathbf{k}),\end{aligned} \tag{5}$$

where $\sim$ denotes the Fourier transform. It should be noted that the electric energy density of the illumination has to be uniform ($U_e(\mathbf{r}) = U_e$) while generating the structured OC distribution so that the second and third term on the right-hand side of Eq. (5) which contain high spatial frequency information ($\pm 1^{st}$ order components) can be extracted. The first term on the right-hand side of Eq. (5) which represents the signal from the electric dipolar absorption ($0^{th}$ order component) will be discarded during chiral SIM image reconstruction (Supplemental Material). After recombining the $\pm 1^{st}$ order components with appropriate weighting factors, the chiral domain image with sub-diffraction limited resolution can be reconstructed.

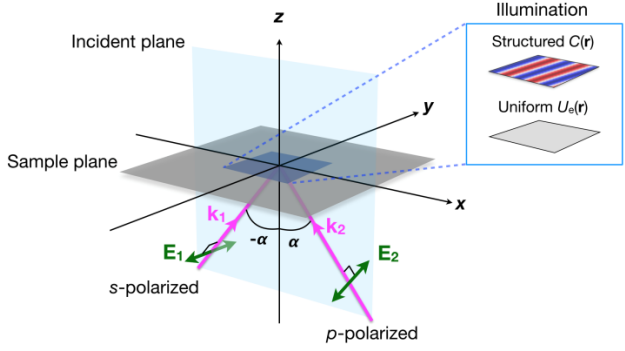

FIG. 2. Schematic of a possible approach to generate structured OC patterns by far-field optics. The superposition of *s*- and *p*-polarized beams with an incident angle of $\pm\alpha$ forms periodically structured OC pattern $C(\mathbf{r})$ and uniform electric energy density distribution $U_e(\mathbf{r})$ on the sample plane.

In the following, we outline one possible experimental scheme to produce structured OC patterns shown in Fig. 1(b) by far-field optics and theoretically demonstrate chiral SIM. As depicted in Fig. 2, the superposition of *s*- and *p*-polarized beams with the incident angle of $\pm\alpha$ forms periodically structured OC pattern and uniform electric energy density distribution on the sample plane (inset of Fig. 2). A Siemens star made of left-handed polyfluorene film embedded in its enantiomer is simulated as a chiral sample. Annealed polyfluorene film is known to exhibit a relatively large chiral response under excitation around 400 nm [39]. The magnitude of the averaged dissymmetry factor of the film goes up to 0.37 [11], which provides sufficient fluorescence modulation for the image reconstruction. For chiral samples with weak CD, the noise becomes relatively pronounced and enhancement in OC might be necessary (Supplemental Material).

To mimic a real experiment, the finite-difference time-domain method (FDTD Solutions, Lumerical) has been applied to simulate the structured OC patterns (Supplemental Material). In the simulation, two plane waves ($\lambda = 405$ nm) with *k*-vectors in the incident plane (*xz*-plane, $\alpha = 57°$) are superimposed to obtain the structured OC patterns. The patterns are generated in three in-plane orientations with three modulation phases for isotropic resolution improvement. As a reference for benchmarking the chiral SIM method, a conventional wide-field FDCD image is acquired under the illumination of left- and right-handed CPLs.

Figure 3 presents the theoretical demonstration of chiral SIM. All of the images are normalized to their largest pixel value and the maximum received photon number of each pixel is set to be $10^5$. The simulated wide-field FDCD and chiral SIM images of the Siemens star are shown in Figs 3(a) and (b), respectively. As can be seen, the chiral SIM image has a reduced blurry central area, indicating an improved spatial resolution due to the expanded *k*-space

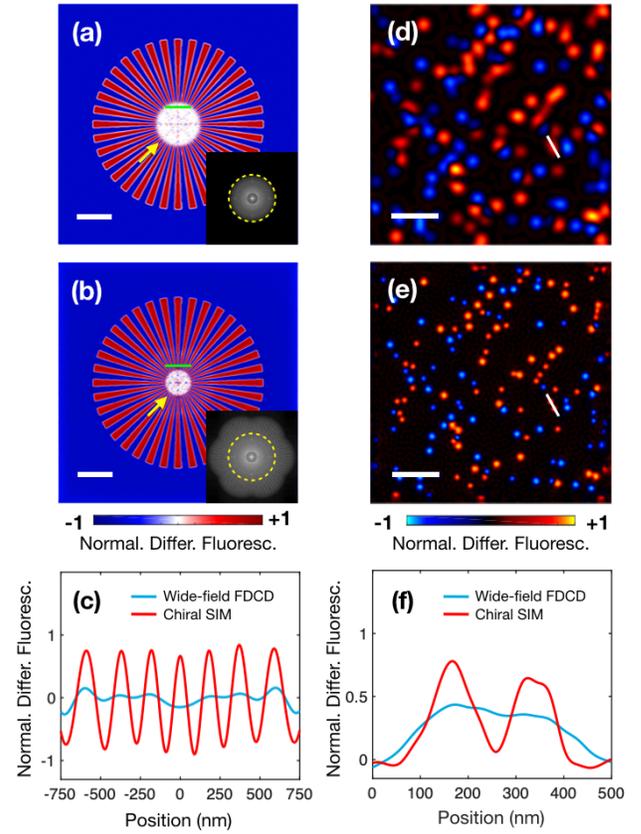

FIG. 3. Theoretical demonstration of chiral SIM. (a) Simulated wide-field FDCD image and (b) simulated chiral SIM image of a left-handed Siemens star embedded in its enantiomer surrounded by a background. The arrows mark the onset of the unresolved blurry area due to the limited spatial resolution. The insets show the corresponding Fourier domain images with dashed circles indicating the optical transfer function. Scale bar: 2 µm. (c) Spatial profiles of the wide-field FDCD image (blue) and chiral SIM image (red) along the green lines in (a) and (b). (d) Simulated wide-field FDCD image and (e) simulated chiral SIM image of randomly distributed chiral beads in an achiral background. Scale bar: 1 µm. (f) Spatial profiles of the wide-field FDCD image (blue) and chiral SIM image (red) along the white lines in (d) and (e). Color bars indicate normalized differential fluorescence.

bandwidth beyond the optical transfer function of the optical system. The boundaries between opposite chiral domains manifest themselves as pseudo achiral domains. This is a universal feature of all CD images due to the finite spatial resolution, regardless of the imaging methods [8,15,17,19]. From the line-cut profiles in Fig. 3(c), the boundaries between opposite chiral domains can be better resolved with the chiral SIM method. In Figs. 3(d-f), randomly distributed nanobeads (diameter of 100 or 150 nm) made of the enantiomers of a chiral material (dissymmetry factor = 0.37) are simulated to mimic a realistic sample. Compared to the wide-field FDCD image (Fig. 3(d)), the chiral SIM image (Fig. 3(e)) shows a higher resolution and, therefore, higher intensity at the location of the beads. Unresolved groups of beads in the wide-field

FDCD image are now clearly resolved as separated spots in the chiral SIM image, as evident in the line-cut profiles shown in Fig. 3(f).

Finally, we address the theoretical spatial resolution of chiral SIM using far-field OC engineering approach shown in Fig. 2. When the same objective is used for excitation and collection, the resolution can be expressed as

$$\delta = \frac{2\pi}{k_{\text{cutoff}} + |\mathbf{k}_C|}, \qquad (6)$$

where $k_{\text{cutoff}} = 4\pi \text{NA}/\lambda_{\text{em}}$ is the cutoff frequency of the imaging system with NA being the numerical aperture of the objective and $\lambda_{\text{em}}$ the emission wavelength. For the OC pattern generated by a pair of *s*- and *p*-polarized beams (Fig. 2), the maximum $|\mathbf{k}_C|$ is $4\pi \text{NA}/\lambda_{\text{ex}}$ with $\lambda_{\text{ex}}$ being the excitation wavelength (Supplemental Material). Assuming $\lambda_{\text{em}}$ is very close to $\lambda_{\text{ex}}$, the highest spatial resolution improvement of chiral SIM over wide-field FDCD imaging method is $(k_{\text{cutoff}} + |\mathbf{k}_C|)/k_{\text{cutoff}} \sim 2$ which is the same as that of typical SIM. In our theoretical demonstration, $|\mathbf{k}_C|/k_{\text{cutoff}}$ is set to be 0.7. The final resolution of chiral SIM is around 100 nm with $\lambda_{\text{ex}} = 405$ nm and NA = 1.2. To further improve the spatial resolution, one may exploit plasmonic SIM scheme [40,41] and generate OC patterns using the optical near fields of the resonant plasmonic nanostructures [34-36]. The proposed chiral SIM method may find applications in sub-cellular bio-imaging [7,15,42] and material sciences [8,43].

The support from the DFG (Receptor Light TRR 166 TP BO6, HU2626/3-1 and CRC 1375 NOA) and the innovation project of Leibniz Institute of Photonic Technology are acknowledged. J.Z. acknowledges the support from 2018 Sino-German (CSC-DAAD) Postdoc Scholarship Program. The authors declare the following competing financial interest: J.-S.H. has filed a patent application for chiral SIM.

[1] N. Berova, K. Nakanishi, and R. W. Woody, *Circular dichroism: principles and applications* (John Wiley & Sons, New York, 2000).
[2] K. W. Busch and M. A. Busch, *Chiral analysis* (Elsevier, Amsterdam, 2011).
[3] N. J. Greenfield, Nat. Protoc. **1**, 2876 (2006).
[4] K. Matsuo, Y. Sakurada, R. Yonehara, M. Kataoka, and K. Gekko, Biophys. J. **92**, 4088 (2007).
[5] A. Micsonai, F. Wien, L. Kernya, Y.-H. Lee, Y. Goto, M. Réfrégiers, and J. Kardos, Proc. Natl. Acad. Sci. U.S.A. **112**, E3095 (2015).
[6] A. J. Miles and B. A. Wallace, Chem. Soc. Rev. **45**, 4859 (2016).
[7] F. Livolant and M. F. Maestre, Biochemistry **27**, 3056 (1988).
[8] K. Claborn, E. Puklin-Faucher, M. Kurimoto, W. Kaminsky, and B. Kahr, J. Am. Chem. Soc. **125**, 14825 (2003).
[9] O. Arteaga, M. Baldrís, J. Antó, A. Canillas, E. Pascual, and E. Bertran, Appl. Opt. **53**, 2236 (2014).
[10] W. Mickols and M. F. Maestre, Rev. Sci. Instrum. **59**, 867 (1988).
[11] M. Savoini, P. Biagioni, S. C. Meskers, L. Duo, B. Hecht, and M. Finazzi, J. Phys. Chem. Lett. **2**, 1359 (2011).
[12] L. M. Haupert and G. J. Simpson, Annu. Rev. Phys. Chem. **60**, 345 (2009).
[13] X. Chen, O. Nadiarynkh, S. Plotnikov, and P. J. Campagnola, Nat. Protoc. **7**, 654 (2012).
[14] N. Mazumder, J. Qiu, M. R. Foreman, C. M. Romero, C.-W. Hu, H.-R. Tsai, P. Török, and F.-J. Kao, Opt. Express **20**, 14090 (2012).
[15] H. Lee, M. J. Huttunen, K.-J. Hsu, M. Partanen, G.-Y. Zhuo, M. Kauranen, and S.-W. Chu, Biomed. Opt. Express **4**, 909 (2013).
[16] G.-Y. Zhuo, H. Lee, K.-J. Hsu, M. Huttunen, M. Kauranen, Y.-Y. Lin, and S.-W. Chu, J. Microsc. **253**, 183 (2014).
[17] S. P. Rodrigues, S. Lan, L. Kang, Y. Cui, and W. Cai, Adv. Mater. **26**, 6157 (2014).
[18] K. R. Campbell and P. J. Campagnola, J. Phys. Chem. B **121**, 1749 (2017).
[19] M. Savoini, X. Wu, M. Celebrano, J. Ziegler, P. Biagioni, S. C. Meskers, L. Duò, B. Hecht, and M. Finazzi, J. Am. Chem. Soc. **134**, 5832 (2012).
[20] K. Mawatari, S. Kubota, and T. Kitamori, Anal. Bioanal. Chem. **391**, 2521 (2008).
[21] M. Liu and M. Franko, Crit. Rev. Anal. Chem. **44**, 328 (2014).
[22] M. Savoini, P. Biagioni, G. Lakhwani, S. Meskers, L. Duo, and M. Finazzi, Opt. Lett. **34**, 3571 (2009).
[23] F. Tantussi, F. Fuso, M. Allegrini, N. Micali, I. G. Occhiuto, L. M. Scolaro, and S. Patanè, Nanoscale **6**, 10874 (2014).
[24] Y. Nishiyama and H. Okamoto, J. Phys. Chem. C **120**, 28157 (2016).
[25] Y. Tang and A. E. Cohen, Phys. Rev. Lett. **104**, 163901 (2010).
[26] D. H. Turner, I. Tinoco Jr, and M. Maestre, J. Am. Chem. Soc. **96**, 4340 (1974).
[27] B. Ehrenberg and I. Steinberg, J. Am. Chem. Soc. **98**, 1293 (1976).
[28] I. Tinoco Jr and D. H. Turner, J. Am. Chem. Soc. **98**, 6453 (1976).
[29] R. Heintzmann and C. G. Cremer, SPIE Proc. **3568**, 185 (1999).
[30] M. G. Gustafsson, J. Microsc. **198**, 82 (2000).
[31] A. Jost and R. Heintzmann, Annu. Rev. Mater. Res. **43**, 261 (2013).
[32] E. Ingerman, R. London, R. Heintzmann, and M. Gustafsson, J. Microsc. **273**, 3 (2019).
[33] Y. Tang and A. E. Cohen, Science **332**, 333 (2011).

[34] M. Schäferling, X. Yin, and H. Giessen, Opt. Express **20**, 26326 (2012).
[35] D. Lin and J.-S. Huang, Opt. Express **22**, 7434 (2014).
[36] M. L. Tseng, Z. H. Lin, H. Y. Kuo, T. T. Huang, Y. T. Huang, T. L. Chung, C. H. Chu, J. S. Huang, and D. P. Tsai, Adv. Opt. Mater. **7**, 1900617 (2019).
[37] H. Hu, Q. Gan, and Q. Zhan, Phys. Rev. Lett. **122**, 223901 (2019).
[38] K. Wicker and R. Heintzmann, Nat. Photonics **8**, 342 (2014).
[39] M. Oda, H. G. Nothofer, G. Lieser, U. Scherf, S. Meskers, and D. Neher, Adv. Mater. **12**, 362 (2000).
[40] J. L. Ponsetto *et al.*, ACS nano **11**, 5344 (2017).
[41] J. T. Collins, C. Kuppe, D. C. Hooper, C. Sibilia, M. Centini, and V. K. Valev, Adv. Opt. Mater. **5**, 1700182 (2017).
[42] S. H. Choi *et al.*, Nature **515**, 274 (2014).
[43] W. Ma, L. Xu, A. F. de Moura, X. Wu, H. Kuang, C. Xu, and N. A. Kotov, Chem. Rev. **117**, 8041 (2017).

# Chiral Structured Illumination Microscopy

# – Supplemental Material

Shiang-Yu Huang,[1] Jiwei Zhang,[1] Christian Karras,[1] Ronny Förster,[1]

Rainer Heintzmann[1,2,3,*] and Jer-Shing Huang[1,2,4,5,*]

[1]*Leibniz Institute of Photonic Technology, Albert-Einstein Straße 9, 07745 Jena, Germany*

[2]*Abbe Center of Photonics, Friedrich-Schiller University Jena, Jena, Germany*

[3]*Institute of Physical Chemistry, Friedrich-Schiller University Jena, Jena, Germany*

[4]*Research Center for Applied Sciences, Academia Sinica, 128 Sec. 2, Academia Road, Nankang District, 11529 Taipei, Taiwan*

[5]*Department of Electrophysics, National Chiao Tung University, 1001 University Road, 30010 Hsinchu, Taiwan*

*e-mail address: heintzmann@gmail.com; jer-shing.huang@leibniz-ipht.de*

# S.1 Operational Procedure of Chiral SIM

The operational procedure of chiral SIM is depicted in Fig. S1. In Step 1, the superposition of *s*- and *p*-polarized beams produces a periodically structured OC pattern $C(\mathbf{r}) = C_0 \cos(\mathbf{k}_c \cdot \mathbf{r} + \theta)$ in one specific in-plane orientation with uniform electric energy density distribution $U_e$ on the sample plane. Here, $C_0$ is the amplitude, $\mathbf{k}_c$ the *k*-vector and $\theta$ the spatial phase of the OC pattern. Similar to typical SIM, the structured OC pattern of the illumination is applied in three in-plane orientations, each with three different $\theta$, and then totally nine raw images are produced. In an experiment, the phase $\theta$ can be adjusted by controlling the phase difference between the two incident beams and the in-plan orientation can be selected by rotating the grating image on the spatial light modulator. In step 2, the raw images are Fourier transformed to their *k*-space images and processed to extract the high-frequency components. The $\pm 1^{st}$ order components are retained because they possess the spatial frequency information about the chiral domains we required (the second and third term in Eq. (5) in the main text). The $0^{th}$ order component is discarded since it only contains the information about fluorescence from the electric dipolar absorption (the first term in Eq. (5) in the main text). Besides, the way to determine the absolute argument (i.e., the phase of the Fourier image that describes the handedness of the sample) should be reconsidered. In typical SIM image reconstruction, the arguments of $\pm 1^{st}$ order components are determined by comparing to the argument of the $0^{th}$ order component at the zero point in the Fourier space ($k = 0$). In chiral SIM, however, the $0^{th}$ order component is discarded. Therefore, an extra wide-field FDCD image of the chiral sample must be acquired to serve this purpose. Finally, all $\pm 1^{st}$ order components in three orientations are recombined with specific weighting factors and the reconstructed image can be obtained with inverse Fourier transform (IFT).

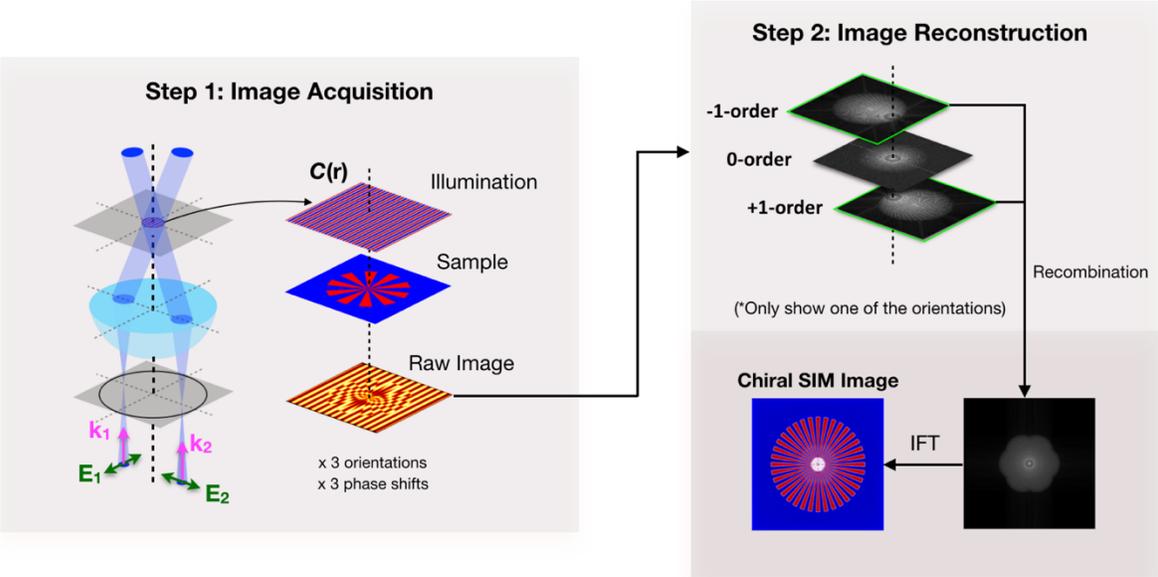

FIG. S1. Illustration of the operational procedure for chiral SIM using far-field optics.

## S.2 Consideration on the Signal Modulation Relative to Noise

Similar to typical SIM, chiral SIM also requires the fluorescence modulation to be sufficiently large compared to the image noise. From Eqs. (2) and (3) in the main text, the fluorescence from a chiral domain with background noise $\sigma$ can be expressed as

$$F + \sigma = \frac{2\beta}{\varepsilon_0}(\omega U_e \alpha'' - CG'') + \sigma. \tag{S.1}$$

The fluorescence modulation comes from the second term, which describes the CD and is determined both by the chirality of the material ($G''$) and the OC ($C$) of the excitation field. As for noise, typical sources associated with imaging cameras are the shot noise $\sigma_S$, dark noise $\sigma_D$ and readout noise $\sigma_R$. Under common imaging conditions and fluorophore brightness, the shot noise is dominant, i.e., $\sigma = \sqrt{\sigma_D^2 + \sigma_R^2 + \sigma_S^2} \approx \sigma_S$. Therefore, the noise depends on the total received photons and thus links to both terms. For chiral samples with large CD, e.g., the chiral polyfluorene film shown in the main text, the fluorescence modulation is large enough for image reconstruction. However, for chiral samples with small CD, e.g., the chiral fluorophore in Ref. 33, an enhancement in the second term of Eq. (S1) by optical field engineering is needed. In this section, we address this issue and comment on the feasibility of chiral SIM method by considering the ratio $R$ of the CD-dependent fluorescence modulation to the shot noise $\sigma_S$, i.e., $R = \left(\frac{4\beta}{\varepsilon_0} CG''\right)/\sigma_S$.

In conventional CD measurement, a chiral sample is successively illuminated by left- (+) and right-handed (−) CPL, resulting in two absorption signals. In this way, the molecular chirality can be described by a dimensionless "dissymmetry factor"

$$g_{\text{CPL}} \equiv 2\frac{A_+ - A_-}{A_+ + A_-} = -\frac{2C_{\text{CPL}}G''}{\omega U_{e,\text{CPL}}\alpha''}. \tag{S.2}$$

For FDCD measurement, the CD-dependent fluorescence is acquired with shot noise $\sigma_S$, which is proportional to the square root of the total received photon number $N_\pm$. Therefore, the ratio $R$ between the fluorescence modulation and noise can be expressed as

$$R = \frac{\frac{4\beta}{\varepsilon_0}CG''}{\sigma_S} = \frac{|N_+ - N_-|}{\sqrt{N_+ + N_-}} = \frac{1}{2}|g_{\text{CPL}}|\sqrt{N_+ + N_-}. \tag{S.3}$$

For chiral sample with very small $g_{\text{CPL}}$, $R$ can be improved by exploiting engineered illumination with enhanced OC, e.g. the optical near fields around well designed plasmonic nanostructures [34,35,36]. In this case, the OC can be larger than that of CPL by a factor of $m$, leading to enhanced CD and thus higher dissymmetry factor $g = mg_{\text{CPL}}$. Therefore, the corresponding $R$ becomes $\frac{1}{2}m|g_{\text{CPL}}|\sqrt{N_+ + N_-}$. In typical SIM, a reasonable signal-to-noise ratio for successful image reconstruction is commonly larger than 10. Therefore, we plot in Fig. S2 the necessary enhancement factor $m$ with respect to $|g_{\text{CPL}}|$ for chiral samples to achieve $R=10$ at two realistic total received photon numbers, namely $N_+ + N_- = 2 \times 10^5$ and $2 \times 10^4$. Figure S2 shows that samples with smaller $|g_{\text{CPL}}|$ needs either larger enhancement factor or higher received photon number to achieve $R$

= 10. While the enhancement can be obtained by optical engineering using rationally designed nanostructures, the maximum received photon number is determined by multiple factors, including the saturation level of the CCD, the emission brightness of the sample and the allowed acquisition time.

In the main text, the chiral sample has a $|g_{CPL}| = 0.37$. Therefore, no enhancement of the chiral response is necessary, i.e., $m = 1$. However, if a chiral sample with weak CD is used, e.g. the chiral fluorophore film in Ref. 33 with $|g_{CPL}| = 1.41 \times 10^{-3}$, an enhancement in the OC is necessary to increase $R$ for successful image reconstruction. Figures S3 (a) and (b) show the reconstructed image of the two samples with $|g_{CPL}| = 0.37$ and $|g_{CPL}| = 1.41 \times 10^{-3}$, respectively, without enhancement ($m = 1$). Obviously, the signal modulation of the latter is insufficient for successful image reconstruction. By increasing the enhancement factor $m$, the quality of the reconstructed images is clearly improved (Figs. S3 (c) and (d)). Overall, the feasibility of chiral SIM depends on the chiroptical response of the material, the OC of the illumination and the experimental conditions. For a chiral sample with small $|g_{CPL}|$, using illumination with enhanced OC or longer acquisition time are both possible ways to increase the quality of the reconstructed image. The latter also requires the consideration on the allowed acquisition time (depending on the sample) and the performance (sensitivity and saturation level) of the CCD.

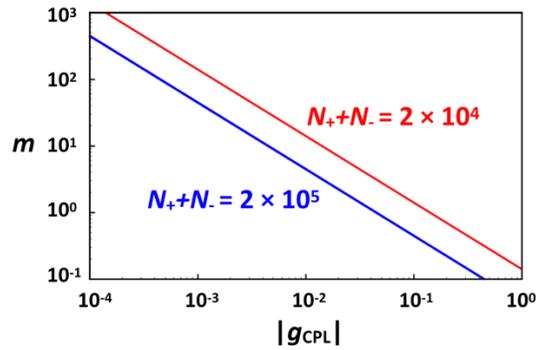

FIG. S2. The enhancement factor $m$ versus dissymmetry factor $|g_{CPL}|$ when $R = 10$ at $N_+ + N_- = 2 \times 10^4$ (red) and $2 \times 10^5$ (blue).

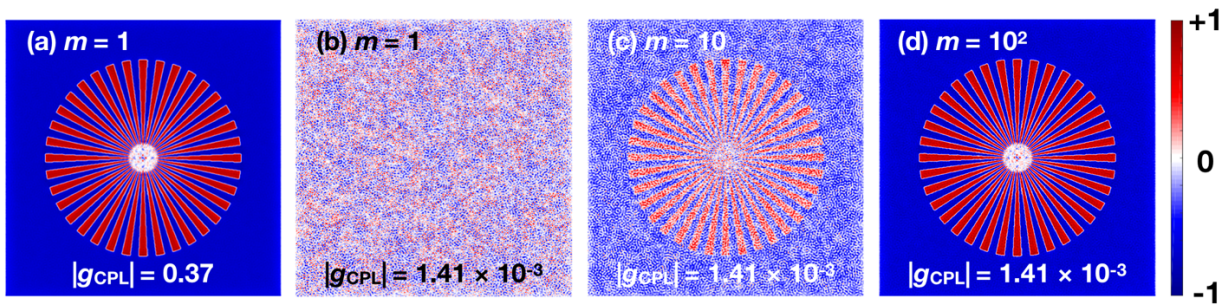

FIG. S3. Theoretical demonstration of the effect of dissymmetry factor $|g_{CPL}|$ and enhancement factor $m$ on the reconstructed chiral SIM image. (a) Chiral SIM image of a sample with $|g_{CPL}| = 1.41 \times 10^{-3}$ and $m = 1$. (b-d) Chiral SIM images of a sample with $|g_{CPL}| = 0.37$ and $m = 1$, $m = 10$ and $m = 10^2$, correspondingly. Color bar indicates normalized differential fluorescence.

## S.3 FDTD Simulation for Structured OC

To simulate the OC patterns produced from the superposition of a pair of *s*- and *p*-polarized lights, we employ commercially available finite-difference time-domain (FDTD) solver (FDTD Solutions, Lumerical). Two plane waves ($\lambda = 405$ nm) are propagating along *z*-axis with $\alpha = 57°$ (Fig. 2 in the main text). A 2D monitor is placed on *xy* plane to record the optical field for further calculation of the OC pattern. The simulation region is set to be 15 μm × 15 μm × 0.7 μm with Bloch boundaries in *x* and *y* directions and perfectly matched layer boundaries in *z* direction. Spatial discretization of 25 nm in *x*, *y* and *z* directions has been applied to a central mesh override region of 15 μm × 15 μm × 0.1 μm. The simulated OC patterns are shown in Fig. S4. The phases of the structured OC patterns are shifted by changing the initial phase of one of the light sources. For acquiring the OC patterns with different orientations, the incident plane waves are rotated around the *z*-axis at three angles (0°, 60° and 120°) while keeping the two waves propagating in the corresponding directions.

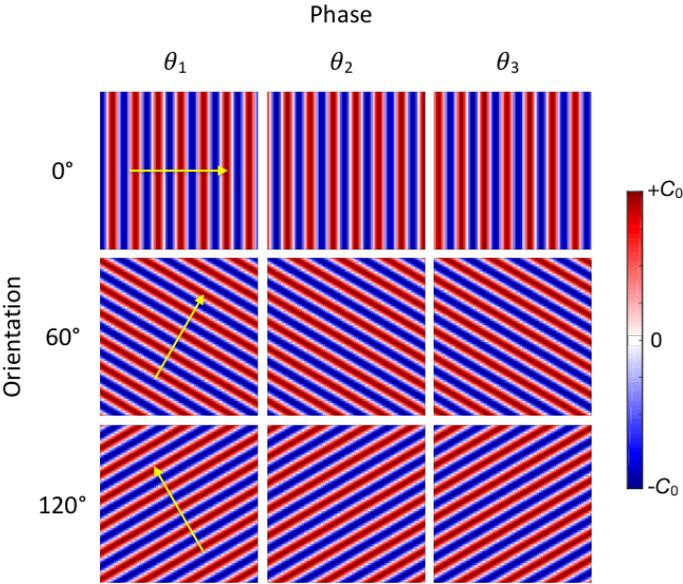

FIG. S4. FDTD-simulated structured OC patterns generated by the superposition of a pair of *s*- and *p*-polarized lights. $C_0$ denotes the amplitude of the structured OC patterns.

## S.4 Maximum Spatial Frequency of OC Pattern

As shown in Fig. 2 in the main text, the electric field of the *s*-polarized plane wave propagating along *z*-axis with an incident angle $\alpha$ on the *xz* plane can be expressed as

$$\mathbf{E}_1 = E_0 e^{ink_0(x\sin\alpha + z\cos\alpha)} e^{-i(\omega t - \varphi_1)} (0,1,0)^T, \tag{S.3}$$

where $E_0$ and $\varphi_1$ are the amplitude and initial phase of the electric field respectively, $n$ is the refractive index of sample environment, $k_0$ and $\omega$ are the wavenumber and angular frequency in vacuum respectively. Accordingly, the magnetic field is described by

$$\mathbf{B}_1 = \frac{nk_0}{\omega} E_0 e^{ink_0(x\sin\alpha + z\cos\alpha)} e^{-i(\omega t - \varphi_1)} (-\cos\alpha, 0, \sin\alpha)^T, \tag{S.4}$$

Similarly, the electric and magnetic field of the *p*-polarized plane wave propagating in another direction can be obtained as

$$\mathbf{E}_2 = E_0 e^{ink_0(-x\sin\alpha + z\cos\alpha)} e^{-i(\omega t - \varphi_2)} (\cos\alpha, 0, \sin\alpha)^T, \tag{S.5}$$

$$\mathbf{B}_2 = \frac{nk_0}{\omega} E_0 e^{ink_0(-x\sin\alpha + z\cos\alpha)} e^{-i(\omega t - \varphi_2)} (0,1,0)^T, \tag{S.6}$$

where $\varphi_2$ is the initial phase of the *p*-polarized plane wave. Therefore, the superposition of these two plane waves is

$$\mathbf{E} = E_0 e^{ink_0 z \cos\alpha} e^{-i\omega t} \left[ \cos\alpha\, e^{i(-nk_0 x \sin\alpha + \varphi_2)}, e^{i(nk_0 x \sin\alpha + \varphi_1)}, \sin\alpha\, e^{i(-nk_0 x \sin\alpha + \varphi_2)} \right]^T, \tag{S.7}$$

$$\mathbf{B} = \frac{nk_0}{\omega} E_0 e^{ink_0 z \cos\alpha} e^{-i\omega t} \left[ -\cos\alpha\, e^{i(nk_0 x \sin\alpha + \varphi_1)}, e^{i(-nk_0 x \sin\alpha + \varphi_2)}, \sin\alpha\, e^{i(nk_0 x \sin\alpha + \varphi_1)} \right]^T, \tag{S.8}$$

Finally, the resulted OC is calculated by

$$C = -\frac{\varepsilon_0 \omega}{2} \text{Im}(\mathbf{E}^* \cdot \mathbf{B}) = \varepsilon_0 E_0^2 n k_0 \cos^2\alpha \sin\Phi, \tag{S.9}$$

where, $\Phi = 2nk_0 x \sin\alpha + \Delta\varphi$, $\Delta\varphi = \varphi_1 - \varphi_2$ is the phase difference of the plane waves. From Eq. (S.9), the generated OC is structured because of the existing $\sin\Phi$ term and the absolute value of wave vector of OC pattern is $k_C = |\mathbf{k}_C| = 2nk_0 \sin\alpha$. Considering $k_0 = \frac{2\pi}{\lambda_{\text{ex}}}$ and the numerical aperture of the imaging system $\text{NA} = n\sin\alpha_{\max}$, the maximum of $k_C$ is determined by $k_{C,\max} = \frac{4\pi \text{NA}}{\lambda_{\text{ex}}}$.